\documentstyle[11pt]{article}
\newcommand{\qab}{{q_{\alpha \beta}}}

\newcommand{\sumab}{{\sum_{\alpha \beta}}}
\newcommand{\suma}{{\sum_\alpha}}

\newcommand{\hatq}{\hat{q}}
\newcommand{\qhab}{{ \hatq_{\alpha \beta}}}

\newcommand{\vsp}{\vspace*{3mm}}
\newcommand{\be}{\begin{equation}}
\newcommand{\ee}{\end{equation}}
\newcommand{\bd}{\begin{displaymath}}
\newcommand{\ed}{\end{displaymath}}
\newcommand{\bea}{\begin{eqnarray}}
\newcommand{\eea}{\end{eqnarray}}

\newcommand{\extr}{~{\rm extr}}
\newcommand{\bra}{\langle}
\newcommand{\ket}{\rangle}
\newcommand{\bigbra}{\left\langle}
\newcommand{\bigket}{\right\rangle}
\newcommand{\order}{{\cal O}}
\newcommand{\minus}{\!-\!}
\newcommand{\plus}{\!+\!}

\newcommand{\erf}{{\rm erf}}
\newcommand{\Size}{260}
\newcommand{\BSize}{200}
\newcommand{\one}{1\!\!{\rm I}}

\newcommand{\room}{\rule[-0.3cm]{0cm}{0.8cm}}
\newcommand{\bigroom}{\rule[-0.5cm]{0cm}{1.2cm}}

\newcommand{\bsigma}{{\mbox{\boldmath $\sigma$}}}

\newcommand{\bxi}{{\mbox{\boldmath $\xi$}}}
\newcommand{\bLambda}{{\mbox{\boldmath $\Lambda$}}}

\newcommand{\bm}{\mbox{\boldmath $m$}}

\newcommand{\hq}{{\hat{q}}}
\newcommand{\hbq}{\hat{\mbox{\boldmath $q$}}}
\newcommand{\bq}{\mbox{\boldmath $q$}} 
\newcommand{\bz}{\mbox{\boldmath $z$}} 

\begin{document}
\title{\bf Feed-Forward Chains of Recurrent Attractor Neural Networks
Near Saturation}
\author{A.C.C. Coolen \dag \and L. Viana \ddag}
\date{June 26, 1996}
\maketitle
 
\begin{center}
\dag~ Dept. of Mathematics, King's College, University of London\\
Strand, London WC2R 2LS, U.K.
\end{center}
\begin{center}
\ddag ~Lab. de Ensenada, Instituto de F\'{\i}sica, UNAM\\
A. Postal 2681, 22800 Ensenada, B.C., M\'{e}xico
\end{center} 
\vspace*{0mm}

\begin{center}
\end{center}
 
\begin{abstract}
\noindent
We perform a stationary state replica analysis for a layered
network of Ising spin neurons, with recurrent Hebbian 
interactions within each layer, in combination with
strictly feed-forward Hebbian interactions between successive layers. 
This model interpolates between the fully 
recurrent and symmetric attractor  
network studied by Amit el al, and the strictly feed-forward attractor network 
studied by Domany et al.
Due to the absence of detailed balance, it is as yet solvable
only in the zero temperature limit.   
The built-in competition between two  
qualitatively different modes of operation, feed-forward (ergodic
within layers)   
versus recurrent (non-ergodic within layers), is found to induce interesting phase transitions. 
\end{abstract}

\tableofcontents 

\pagebreak
\section{Introduction}
There are two main classes of solvable models for neural networks,
functioning as associative information processing devices.  
The first class consists of models in which the matrix of interactions between
the neurons is symmetric \cite{hopfield,amitetal1,amitetal2}.
This ensures detailed balance, so that equilibrium
statistical mechanics applies.  The
second class of models are equipped with neuron interactions
which are non-symmetric, but are of a strictly feed-forward
nature.  Such models can be constructed explicitly,
by arranging the neurons
in feed-forward layers \cite{domanyetal1,domanyetal2,domanyetal3},
or implicitly, by introducing an
extreme dilution such that on finite time-scales
the systems behave effectively tree-like \cite{DGZ}.  The latter models
can no longer be studied within equilibrium statistical
mechanics; here it is the strict feed-forward nature of
the interactions which enables a dynamical solution. 
Even the more recent studies of layered attractor networks all rely 
for solution 
on the connectivity being either strictly symmetric 
(e.g. \cite{BAM1,BAM2}) or strictly feed-forward (e.g.
\cite{WCS}).     
 
In this paper we study a dual model, where there is  
a controllable competition between the highly non-ergodic
recurrent information processing, typical for symmetric
systems,
and the feed-forward processing typical for tree-like
structures, where individual layers are ergodic.
We consider chains of equally large neural layers,
each consisting of Ising spin neurons with recurrent interactions, 
in combination with  strictly
feed-forward interactions between succesive layers.
All allowed interactions are of a Hebbian type; this choice
guarantees meaningful information processing, induces familiar
order parameters, and creates convenient benchmarks, in that 
in the two extreme limits of fully recurrent and fully feed-forward
connectivity the familiar results of \cite{amitetal1,amitetal2} and \cite{domanyetal1,domanyetal2,domanyetal3}
must be recovered. 
Each individual layer can be seen as a recurrent attractor 
network with association cues provided by (thermally fluctuating) 
external fields.  
These fields, however, contain contributions from non-nominated patterns,
and therefore contain the disorder variables to be averaged out
(which already appear in the recurrent interactions), in
contrast to the situation with most other attractor models with
external fields, such as \cite{rau1,rau2,engel,YW}. 

Our motivation for studying this type of model is threefold. From a
statistical mechanical point of view the model is of interest simply 
because its interaction matrix is neither symmetric nor strictly feed-forward, so that the 
conventional routes to a solution cannot be followed, and since it
interpolates between two systems with quite different properties. 
Secondly,
because the model can 
nonetheless be solved (at least in the zero temperature limit),   
it could serve as a convenient future playground for 
testing new formalisms aimed at calulating stationary state properties
of non-symmetric neural network models, such as \cite{nobalance}. 
Thirdly, from a biological point of view, both recurrent and
feed-forward information processing have specific advantages and
disadvantages, and especially in the peripherical brain regions one 
therefore usually finds both interaction types present. Recent
experimental evidence even suggests that 
the balance between recurrent 
and feed-forward processing in these regions is actively controlled by
neuromodulators \cite{verschure}. Yet, as far as we
are aware, no dual models have been solved so far.

This paper is organised as follows. After having defined our model, we
analyse its zero temperature stationary state, 
using replica theory. The resulting saddle-point equations are solved
upon making the replica symmetry (RS) ansatz, leaving two 
control parameters: $\alpha$ (the information storage level)
and $\omega$ (a parameter reflecting the balance between recurrent
and feed-forward operation).   We then study the
various types of phase transitions exhibited by the model: $(i)$ the
saturation  (or storage capacity) transition, $(ii)$ simple (RS) ergodicity
breaking transitions, and $(iii)$ complex (i.e. replica symmetry
breaking, RSB) ergodicity-breaking transitions. 
All analytical results are confirmed by numerical simulations.

\section{Model Definition and Solution}
 
\subsection{Model Definition}

\begin{figure}[h]
\vsp
\setlength{\unitlength}{0.07mm}
\begin{center}\begin{picture}(1300,850)(200,50)
\put(220,200){\vector(1,0){\Size}}
\put(220,500){\vector(1,0){\Size}}
\put(220,800){\vector(1,0){\Size}}
\put(220,200){\vector(3,1){\Size}}
\put(220,200){\vector(3,2){\Size}}
\put(220,200){\vector(1,1){\Size}}
\put(220,200){\vector(3,4){\Size}}
\put(220,500){\vector(1,1){\Size}}
\put(220,500){\vector(3,2){\Size}}
\put(220,500){\vector(3,1){\Size}}
\put(220,500){\vector(3,-1){\Size}}
\put(220,500){\vector(3,-2){\Size}}
\put(220,500){\vector(1,-1){\Size}}
\put(220,800){\vector(3,-1){\Size}}
\put(220,800){\vector(3,-2){\Size}}
\put(220,800){\vector(1,-1){\Size}}
\put(220,800){\vector(3,-4){\Size}}
\put(600,500){\vector(3,2){80}}
\put(600,500){\vector(3,-2){80}}
\put(600,500){\vector(-3,2){80}}
\put(600,500){\vector(-3,-2){80}}
\put(500,150){\framebox(\BSize,700)}
\put(480,75){Layer $\ell\!\!-\!\!1$}
\put(540,890){$\bsigma^{\ell-1}$}
\put(510,600){$\{J_{ij}^{\ell-1}\}$}
\put(800,850){$\{W_{ij}^\ell\}$}
\put(720,200){\vector(1,0){\Size}}
\put(720,500){\vector(1,0){\Size}}
\put(720,800){\vector(1,0){\Size}}
\put(720,200){\vector(3,1){\Size}}
\put(720,200){\vector(3,2){\Size}}
\put(720,200){\vector(1,1){\Size}}
\put(720,200){\vector(3,4){\Size}}
\put(720,500){\vector(1,1){\Size}}
\put(720,500){\vector(3,2){\Size}}
\put(720,500){\vector(3,1){\Size}}
\put(720,500){\vector(3,-1){\Size}}
\put(720,500){\vector(3,-2){\Size}}
\put(720,500){\vector(1,-1){\Size}}
\put(720,800){\vector(3,-1){\Size}}
\put(720,800){\vector(3,-2){\Size}}
\put(720,800){\vector(1,-1){\Size}}
\put(720,800){\vector(3,-4){\Size}}
\put(1100,500){\vector(3,2){80}}
\put(1100,500){\vector(3,-2){80}}
\put(1100,500){\vector(-3,2){80}}
\put(1100,500){\vector(-3,-2){80}}
\put(1000,150){\framebox(\BSize,700)}
\put(1020,75){Layer $\ell$}
\put(1080,890){$\bsigma^{\ell}$}
\put(1050,600){$\{J_{ij}^{\ell}\}$}
\put(1220,200){\vector(1,0){\Size}}
\put(1220,500){\vector(1,0){\Size}}
\put(1220,800){\vector(1,0){\Size}}
\put(1220,200){\vector(3,1){\Size}}
\put(1220,200){\vector(3,2){\Size}}
\put(1220,200){\vector(1,1){\Size}}
\put(1220,200){\vector(3,4){\Size}}
\put(1220,500){\vector(1,1){\Size}}
\put(1220,500){\vector(3,2){\Size}}
\put(1220,500){\vector(3,1){\Size}}
\put(1220,500){\vector(3,-1){\Size}}
\put(1220,500){\vector(3,-2){\Size}}
\put(1220,500){\vector(1,-1){\Size}}
\put(1220,800){\vector(3,-1){\Size}}
\put(1220,800){\vector(3,-2){\Size}}
\put(1220,800){\vector(1,-1){\Size}}
\put(1220,800){\vector(3,-4){\Size}}
\end{picture}
\end{center}
\vspace*{-2mm}
\caption{Two adjacent modules 
in the $L$-layer chain, with their associated microscopic variables
and parameters: $\bsigma^\ell\in\{-1,1\}^N$ (neuron states in layer
$\ell$), $\{J_{ij}^\ell\}$ (symmetric recurrent interactions in layer
$\ell$), and $\{W_{ij}^\ell\}$ (feed-forward interactions from layer
$\ell\minus 1$ to layer $\ell$).}
\label{fig:architecture} 
\end{figure}
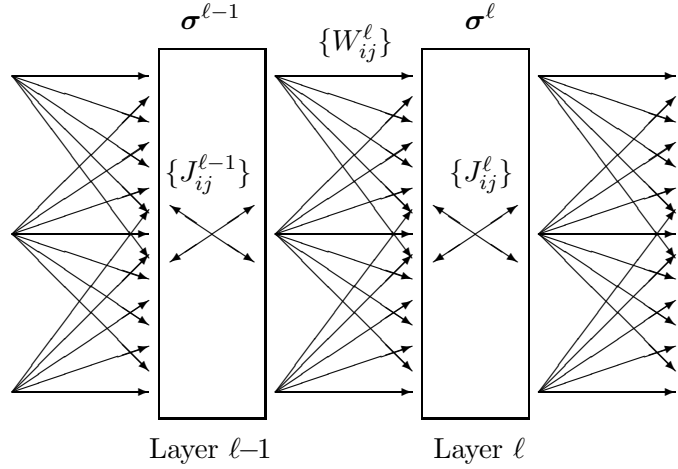

Our model system is a feed-forward chain of $L$ recurrent layers, 
each of which consists of $N$ Ising spin 
neurons. The variable 
$\sigma_i^\ell\in\{\minus 1,1\}$ denotes the state
(non-firing/firing) of neuron $i$ in layer $\ell$, and  
the collective state of the $N$ neurons in layer
$\ell$ is written as
$\bsigma^{\ell}=(\sigma_1^\ell,\ldots,\sigma_N^\ell)$. 
The dynamics is the usual Glauber-type sequential stochastic alignment of the 
neurons $\sigma_i^\ell$ to local fields $h_i^\ell$, to which here only neurons 
from within layer $\ell$ (via recurrent interactions) and neurons from the
previous layer 
$\ell\minus 1$ (via feed-forward
interactions) are allowed to contribute (see figure \ref{fig:architecture}):
\be
{\rm Prob}[\sigma_i^\ell\to \minus \sigma_i^\ell]=\frac{1}{2}[
1\minus \tanh(\beta\sigma_i^\ell h_i^\ell(\bsigma^\ell,\bsigma^{\ell-1}))]
~~~~~~~~
h_i^\ell(\bsigma^\ell,\bsigma^{\ell-1})=
\sum_{j} J_{ij}^\ell \sigma_j^\ell \plus
\sum_j W_{ij}^\ell \sigma_j^{\ell-1}
\label{eq:ratesandfields}
\ee
At each time step the candidate neuron $(i,\ell)$ to be updated
is drawn at random from $\{1,\ldots,N\}\times\{1,\ldots,L\}$. 
The parameter $\beta=T^{-1}$ controls the amount of noise, with $T=0$
corresponding to deterministic alignment (although the order of
updates still remains random).  

\noindent
The $2L\times N^2$ synaptic interactions are defined 
as the result of the network having learned $p=\alpha N$ 
patterns with a Hebbian-type rule:
\be
J_{ij}^\ell=\frac{J_0}{N}[1\minus \delta_{ij}]\sum_{\mu=1}^{p} \xi_i^{\mu,\ell}\xi_j^{\mu,\ell}
~~~~~~~~
W_{ij}^\ell=\frac{J}{N}\sum_{\mu=1}^{p}
\xi_i^{\mu,\ell}\xi_j^{\mu,\ell-1} 
\label{eq:synapses}
\ee
with $\xi_i^{\mu,\ell}\in\{-1,1\}$ denoting component $i$
in layer $\ell$ of pattern $\mu$ ($\mu=1,\ldots,p$). 
Each pattern represents a specific microscopic neural configuration in
the chain as a whole. 
For simplicity all pattern components are
drawn independently at random from $\{\minus 1,1\}$, and $N$ is
assumed to be large (eventually we will take the limit $N\to\infty$). 
The two parameters $(J_0,J)$ (one of which will become redundant in the
limit $T\to 0$) control the 
relative strength of the two interaction types. 
The only exception to (\ref{eq:synapses}) is the first layer, where by
definition we
have to set $W_{ij}^{1}=0$ $(\forall ij)$, and where we have to
distinguish between two modes of operation: $(i)$ free relaxation of
layer 1, following a specific initialisation which serves as the
recall cue, and $(ii)$ so-called 'clamped' operation, where the recall
cue is provided by the state
vector $\bsigma^1$ itself, which is stationary and specified externally.  

For $J=0$ our model reduces to a collection of $L$ decoupled symmetric
attractor networks of the Hopfield \cite{hopfield} type, which can be
solved in equilibrium using equilibrium statistical mechanics
\cite{amitetal1,amitetal2}. Such systems are known to have a storage capacity  
 of $\alpha_c\sim 0.138$ (for $T=0$, in RS approximation),   
and non-trivial ergodicity breaking (RSB) for sufficiently low temperatures. 
For $J_0=0$, on the other hand, we have feed-forward
interactions only. Now the local fields are found to have Gaussian
probability distributions in the stationary state, enabling the 
derivation of recurrent relations for the values of order parameters 
in subsequent layers
\cite{domanyetal1,domanyetal2,domanyetal3}. Here one finds a storage capacity of 
$\alpha_c\sim 0.269$ (for $T=0$) and  no RSB at any temperature. 
The solution of the present model will have to represent a marriage
of these two extremes, such that the equations first derived in
\cite{amitetal1} and \cite{domanyetal1} follow as special cases, 
upon taking the limits $J\to 0$ and $J_0\to 0$, respectively.  

Away from saturation, for $\alpha=p/N\to 0$, the 
dynamics of this model can be solved easily using existing techniques
(for a review see \cite{CSreview}), leading to a set of coupled
differential equations for a small number of macroscopic observables. 
In order to solve the model near saturation (i.e. for $\alpha>0$), we will exploit the fact that, due to the
strictly feed-forward nature of the inter-layer interactions,  at $T=0$
all layers will eventually go to a stationary state;  for each layer $\ell$ 
there will be a stage after which the input from layer $\ell\minus 1$ 
is stationary. This enables an equilibrium statistical mechanical 
replica analysis at least in the $T\to 0$
limit (for $T>0$ one has to take thermal fluctuations in the external
fields into account, which destroy the Boltzmann form of the
stationary states of the individual layers). 
For technical reasons we will take the limit
$T\to 0$ after the limit 
$N\to \infty$, a commonly made step which in the present case,
however, need not be as harmless as for models obeying detailed
balance (it is justified {\em a posteriori} by the
agreement between theory and numerical simulations).

\subsection{Replica Analysis of the Stationary State}

We analyse the stationary state for a given layer $\ell$, upon
assuming stationary inputs from layer $\ell\minus 1$. Given this
assumption, the dynamics (\ref{eq:ratesandfields}) will evolve towards
equilibrium, characterised by
\be
p_\infty(\bsigma^\ell)\sim e^{-\beta H(\bsigma^\ell)}
~~~~~~
H(\bsigma^\ell)=-\frac{1}{2}\sum_{ij}\sigma^\ell_i J_{ij}^\ell
\sigma^\ell_j-\sum_{ij}\sigma_i^\ell W_{ij}^\ell \sigma_j^{\ell-1}
\label{eq:equil}
\ee
The associated thermal averages are written as $\bra \ldots \ket$. 
We introduce a macroscopic description in terms of the so-called 
overlaps between the system state and the stored patterns, 
and make the usual ansatz that in equilibrium only a finite number $k$ of the
patterns are condensed, which (due to permutation symmetry with
respect to pattern indices) we can take to be $\mu=1,\ldots,k$: 
\be
m_{\mu,\ell}(\bsigma^\ell) = \frac{1}{N}\sum_i \xi^{\mu,\ell}_i\sigma_i^\ell
~~~~~~~~
\left\{
\begin{array}{ll}
\mu\in\{1,\ldots,k\}: & \bra m^2_{\mu,\ell}(\bsigma^\ell)\ket=\order(1)\room \\
\mu\in\{k\plus 1,\ldots,p\}: &
\bra m^2_{\mu,\ell}(\bsigma^\ell)\ket =\order(1/N)\room
\end{array}
\right.
\label{eq:overlaps}
\ee
We assume the free energy per spin $f=- \frac{1}{\beta N}\log Z$ of the system described by (\ref{eq:equil})
to be self-averaging for $N\to\infty$ with respect to the realisation of the
non-condensed patterns $\mu>k$ 
(which play the role of 'frozen disorder'), 
so that we can simplify the calculation of  the free energy by
averaging over these patterns. This property can also be rigorously
proven. To simplify the  pattern average we use the identity $\log 
Z=\lim_{n\to 0}\frac{1}{n}[Z^n\minus 1]$, giving 
\be
- \beta f = 
\lim_{n \to 0} \frac{1}{nN}\left[ \langle Z^n \rangle_{\rm patt} - 1 \right]
~~~~~~~~~~
Z = \sum_{\bsigma\in\{-1,1\}^N}\!\! e^{ - \beta H(\bsigma)}
\label{eq:free}
\ee 
Upon writing the pattern overlaps of the previous layer
$\ell\minus 1$ (which are stationary by assumption) as
$\tilde{m}_\mu=\frac{1}{N}\sum_i\xi_i^{\mu,\ell-1}\sigma_i^{\ell-1}$, 
we can write the 
key quantity in (\ref{eq:free}) for integer values of $n$
in the following form:
\be
\bigbra Z^n \bigket_{\rm patt} 
=e^{-\frac{1}{2}\alpha\beta J_0 n}
\!\!
\sum_{\bsigma^1\in\{-1,1\}^N} \cdots\sum_{\bsigma^n\in\{-1,1\}^N}\!\!
\bigbra
e^{\beta N\sum_{\mu=1}^p\sum_{\alpha=1}^n \left[\frac{1}{2}J_0 m_{\mu,\ell}^2(\bsigma^\alpha)+ J m_{\mu,\ell}(\bsigma^\alpha)\tilde{m}_\mu\right] 
}
\bigket_{\rm patt} 
\label{eq:powersofZ}
\ee
The condensed contribution to (\ref{eq:powersofZ})
(the terms $\mu\leq k$ in the summation over pattern indices)
is linearised by a Gaussian transformation:
\bd
e^{\beta N\sum_{\mu\leq k}\sum_{\alpha} \left[\frac{1}{2}J_0 m_\mu^2(\bsigma^\alpha)
+ J \tilde{m}_\mu m_\mu(\bsigma^\alpha)
\right] 
}
~~~~~~~~~~~~~~~~~~~~~~~~~~~~~~~~~~~~~~~~~~~~~~~~~~~~~~~~~~
\ed
\be
=\left[ \frac{\beta J_0 N}{2\pi} \right]^{\frac{nk}{2}}
\int\! \left[ \prod_{\mu\leq k}\prod_\alpha dm^\alpha_\mu\right]
e^{\sum_{\mu\leq\ell}\sum_\alpha \left[ -\frac{1}{2}\beta J_0 N 
(m^\alpha_\mu)^2 +\beta\sum_{i} \sigma^\alpha_i \xi^{\mu,\ell}_i
(J_0 m^\alpha_\mu + J\tilde{m}_\mu) \right]}
\label{eq:firstbracket}
\ee
The uncondensed contribution to (\ref{eq:powersofZ})
(the terms $\mu> k$ in the summation over pattern indices)
is first linearised by a Gaussian transformation and then 
averaged over the frozen disorder, i.e. the pattern components
$\xi_i^{\mu,\ell}$ with $\mu>k$, giving
\bd
\bigbra e^{\beta N\sum_{\mu> k}\sum_{\alpha} \left[\frac{1}{2}J_0 m_\mu^2(\bsigma^\alpha)
+ J \tilde{m}_\mu m_{\mu,\ell}(\bsigma^\alpha)
\right] 
}
\bigket_{\rm patt}
= \prod_{\mu>k}
\int\!D\bz~
e^{\sum_i\log \cosh\sum_\alpha [ \left(\frac{\beta J_0}{N}\right)^{\frac{1}{2}}
 z_\alpha \sigma_i^\alpha+ \beta J 
\tilde{m}_\mu \sigma_i^\alpha ]}
\ed
with the Gaussian measure $D\bz=\prod_\alpha\left[(2\pi)^{-\frac{1}{2}}e^{-\frac{1}{2}
z_\alpha^2}dz_\alpha\right]$. 
Using the property 
$\tilde{m}^2_\mu=\order(N^{-1})$ for $\mu>k$, 
we expand $\log\cosh(x) = \frac{1}{2}x^2\plus\order(x^4)$. 
Inserting an integral representation of unity to isolate  
the spin glass order parameters $q_{\alpha
\beta}={1 \over N} \sum_i \sigma_i^\alpha \sigma_i^\beta$, 
and repeatedly forgetting about terms which are of vanishing order 
either for $n\to 0$
or for $N\to \infty$, then leads to: 
\bd
\bigbra e^{\beta N\sum_{\mu> k}\sum_{\alpha} \left[\frac{1}{2}J_0 m_\mu^2(\bsigma^\alpha)
+ J m_\mu(\bsigma^\alpha)\tilde{m}_\mu\right] 
}
\bigket_{\rm patt}
~~~~~~~~~~~~~~~~~~~~~~~~~~~~~~~~~~~~~~~~~~~~~~~~~~~~~~~~~~~~~~~~~~~~~~~~~~~~~~~~~~~~~~~~
\ed
\bd
= \int\!d\bq
d\hbq~e^{N\sum_{\alpha\beta}q_{\alpha\beta}\left[i\hat{q}_{\alpha\beta}+
\frac{1}{2}(\beta J)^2\sum_{\mu>k}\tilde{m}^2_\mu\right]
-i\sum_{\alpha\beta}\hat{q}_{\alpha\beta}\sum_i\sigma_i^\alpha\sigma_i^\beta}
\prod_{\mu>k}
\int\! D\bz~
e^{\frac{1}{2}\beta J_0 \bz\cdot\bq\bz 
+\sum_{\alpha}
z_\alpha x_\alpha^\mu }
\ed
\be
= 
\int\!d\bq
d\hbq~e^{N\sum_{\alpha\beta}q_{\alpha\beta}\left[i\hat{q}_{\alpha\beta}+
\frac{1}{2}(\beta J)^2\sum_{\mu>k}\tilde{m}^2_\mu\right]
-i\sum_{\alpha\beta}\hat{q}_{\alpha\beta}\sum_i\sigma_i^\alpha\sigma_i^\beta
-\frac{1}{2}p\log\det\bLambda +\frac{1}{2}\sumab\sum_{\mu>k}
x_\alpha^\mu (\bLambda)^{-1}_{\alpha
\beta} x_\beta^\mu}
\label{eq:secondbracket}
\ee
with the abbreviation 
$x_\alpha^\mu = \beta J (\beta J_0 N)^{1/2}
\tilde{m}_\mu \sum_{\beta}q_{\alpha\beta}$, 
and 
where $\bLambda $
is an $n\times n$ matrix with components
$\Lambda_{\alpha \beta} = \delta_{\alpha \beta} - \beta J_0
\qab$. 
We finally substitute the results
(\ref{eq:firstbracket},\ref{eq:secondbracket}) into
equation (\ref{eq:powersofZ}), and perform the remaining summations over 
the spin variables $\{\sigma_i^{\alpha}\}$. 
The free energy per spin (\ref{eq:free}) is subsequently obtained by interchanging
the limits
$N \to \infty$ and $n \to 0$ and by performing the remaining
integral by the method of steepest descent, leading to the final
result:
\bd
f = \frac{1}{2}\alpha J_0-\beta^{-1}\log 2
- \lim_{n \to 0} \frac{1}{\beta n} \extr \left\{ i \sumab \qhab \qab
- \frac{1}{2} \beta J_0 \suma\sum_{\mu\leq k} [m_\mu^\alpha]^2 
- \frac{1}{2} \alpha \log \det \left[ \one \minus\beta J_0 \bq \right]
\right.
\ed
\be
\left.
+\frac{1}{2} \beta^2 J^2 [ \sum_{\mu > k} \tilde{m}^{2}_\mu]
\sum_{\alpha \beta \gamma} \qab
\left[\one\minus \beta J_0 \bq \right]^{-1}_{\beta \gamma} 
+ \sum_{\bxi} \log 
\sum_{\bsigma} e^{\beta
\suma \sigma^\alpha\sum_{\mu\leq k} [J_0 m_\mu^\alpha +
J \tilde{m}_\mu]\xi_\mu - i\sum_{\alpha\beta}\sigma_\alpha\hq_{\alpha\beta}
\sigma_\beta}
\bigroom\right\}
\label{eq:RSB}
\ee
with $\bxi\in\{\minus 1,1\}^k$ and $\bsigma\in\{\minus 1,1\}^n$. 
The extremum in (\ref{eq:RSB}) refers to variation
of the parameters  
$\{\hat{q}_{\alpha\beta},q_{\alpha\beta},m^\alpha_\mu\}$, and is 
defined as 
the analytical continuation for $n\to 0$ of the saddle point 
which for $n\geq 1$ minimises $f$. The physical meaning of the
parameters $m^\alpha_\mu$ can be deduced for instance by taking the
derivative of 
(\ref{eq:free}) with respect to $\tilde{m}_\mu$, which gives
$\lim_{n\to 0}\frac{1}{n}\sum_\alpha m^\alpha_\mu=
\frac{1}{N}\sum_i\xi_i^{\mu,\ell}\bra\sigma_i^\ell\ket=\bra m_{\mu,\ell}(\bsigma^\ell)\ket$.    
 
\subsection{Replica-Symmetric Solution}
 
We have obtained an expression for the free energy per neuron $f$ in
terms of the 
parameters $\qab$, $\qhab$ and $m^\alpha_\mu$.  In order to take the
limit $n\to 0$ in (\ref{eq:RSB}) and 
obtain an explicit solution,  we will make the replica symmetry (RS)
ansatz for the relevant saddle-point, i.e.:
\bea
\qhab & = & \frac{1}{2} i \alpha \beta^2 \left[ R
\delta_{\alpha \beta} + r(1-\delta_{\alpha \beta}) \right] \nonumber \\
\qab & = & \delta_{\alpha
\beta} + q(1 - \delta_{\alpha \beta})
\label{eq:RSansatz} \\
m^\alpha_\mu & = & m_\mu \nonumber
\eea
Substitution of (\ref{eq:RSansatz}) into (\ref{eq:RSB}) and 
linearisation of the exponent involving the spin variables, allows us
to perform the remaining spin summations and take the limit $n\to 0$,
which gives:
\bd
f_{\rm RS} =
\frac{1}{2}\alpha\left[J_0\plus \beta r(1-q)\right] + \frac{1}{2} J_0 \sum_{\mu\leq k}m_\mu^2
+\frac{\alpha}{2 \beta} \left[ \log[1\minus \beta J_0(1\minus q)] - \frac{\beta J_0 q}
{1\minus \beta J_0(1\minus q)} \right] 
\ed
\be
- \frac{1}{2}[\sum_{\mu >\ell} \tilde{m}^2_\mu] \frac{\beta J^2(1\minus q)}{1\minus \beta J_0(1\minus q)} 
- \frac{1}{\beta}
\bigbra \int\! Dz \log 2 \cosh \beta \left[ \sum_{\mu\leq k}\xi_\mu
( J_0 m_\mu\plus J \tilde{m}_\mu) \plus z \sqrt{\alpha r} \right]
\bigket_\bxi
\label{eq:RS}
\ee
where  $Dz=(2\pi)^{-\frac{1}{2}}e^{-\frac{1}{2}z^2} dz$ and $\bra
G[\bxi] \ket_\bxi=2^{-k}\sum_{\bxi\in\{-1,1\}^k}G[\bxi]$. 
Taking the derivative of the free energy with respect to the control parameter
$J_0$ in addition yields the following identity:
\be
\sum_{\mu>k} \bra m_{\mu,\ell}^2(\bsigma^\ell)\ket =
\alpha-\sum_{\mu\leq k}\bra  m_{\mu,\ell}^2(\bsigma^\ell)\ket
-2\frac{\partial f_{\rm RS}}{\partial J_0}
\label{eq:uncondensed}
\ee
By solving the saddle point equations $\partial f_{\rm RS}/ \partial m_\mu
=\partial f_{\rm RS}/ \partial q =\partial f_{\rm RS}/ \partial r=0$, from 
which we can 
eliminate $\sum_{\mu>k}\tilde{m}_\mu^2$ by applying 
(\ref{eq:uncondensed}) to layer $\ell\minus 1$, 
we finally arrive at a set of recurrent equations 
which relate the values
of order parameters 
in subsequent layers. Upon adapting our notation accordingly, 
these recurrent
relations can be written in their most natural form $(\bm,q,r)\to(\bm^\prime,q^\prime,r^\prime)$:
\be
\bm^\prime = \bra \bxi
\int\! Dz~ \tanh\beta\left[\bxi\cdot
(J_0\bm^\prime\plus J\bm) \plus z \sqrt{\alpha r^\prime}\right] 
\ket_{\bxi}
\label{eq:saddlem}
\ee
\be
q^\prime = \bra
\int\! Dz~ \tanh^2\beta \left[\bxi\cdot
(J_0\bm^\prime\plus  J\bm) \plus z \sqrt{\alpha r^\prime} \right]
\ket_{\bxi}
\label{eq:saddleq}
\ee
\be
r^\prime \left[ 1\minus\beta J_0 (1\minus q^\prime) \right]^2 \minus 
J_0^2q^\prime=
\beta^2 J^2 r(1\minus q)^2
\minus J^2q \plus \frac{J^2(1\plus q)}{1\minus \beta J_0(1\minus q)}
\label{eq:saddler}
\ee
The macroscopic state of every layer $\ell$ is now characterised by
its value for the order
parameter set $(\bm,q,r)$, with $\bm=(m_1,\ldots,m_k)$, and with the 
(familiar) physical meaning
$m_\mu= 
\frac{1}{N}\sum_i\xi_i^{\mu,\ell}\bra\sigma_i^\ell\ket$ and 
$q=\frac{1}{N}\sum_i\bra \sigma_i^\ell\ket^2$. 
The only exceptions to (\ref{eq:saddlem}-\ref{eq:saddler}) arise in
the context of the special status of the first layer. 
If this first layer is clamped into a randomly selected configuration
$\bsigma^1$ with a prescibed vector of the condensed overlaps $\bm$,
then (\ref{eq:saddler}) is to be replaced by
\be 
r^\prime \left[ 1\minus\beta J_0 (1\minus q^\prime) \right]^2 \minus
J_0^2q^\prime=J^2
\room
\label{eq:saddlerclamped}
\ee 
If, on the other hand, the first layer is allowed free relaxation towards equilibrium,
then (\ref{eq:saddlem}-\ref{eq:saddler}) does hold, but with the
macroscopic state $(\bm,q,r)$ of the first layer solved from the set 
corresponding to the
$J=0$ situation, where inter-layer interactions are absent:
\be
\bm= \bra \bxi
\int\! Dz~ \tanh\beta\left[J_0\bxi\cdot\bm \plus z \sqrt{\alpha r}\right] 
\ket_{\bxi}
\label{eq:hopsaddlem}
\ee
\be
q= \bra
\int\! Dz~ \tanh^2\beta \left[J_0\bxi\cdot\bm \plus z \sqrt{\alpha r} \right]
\ket_{\bxi}
\label{eq:hopsaddleq}
\ee
\be
r=\frac{q J_0^2}{\left[ 1\minus\beta J_0 (1\minus q) \right]^2}
\label{eq:hopsaddler}
\ee
In the limit $T\to 0$, the  equations (\ref{eq:saddlem}-\ref{eq:hopsaddler}) 
 constitute the solution of our 
model. We will now analyse their consequences  
and validate their predictions with simulation experiments.
We eliminate the parameter redundancy at $T=0$ by putting 
$J_0=\frac{1}{2}[1\plus \omega]$ and $J=\frac{1}{2}[1\minus \omega]$,
with $\omega\in[\minus 1,1]$.

\section{Phase Transitions}
 
\subsection{Saturation Transition in Infinitely Long Chains}

\begin{figure}[t]
\centering
\vspace*{88mm}
\hbox to \hsize{\hspace*{-10mm}\includegraphics{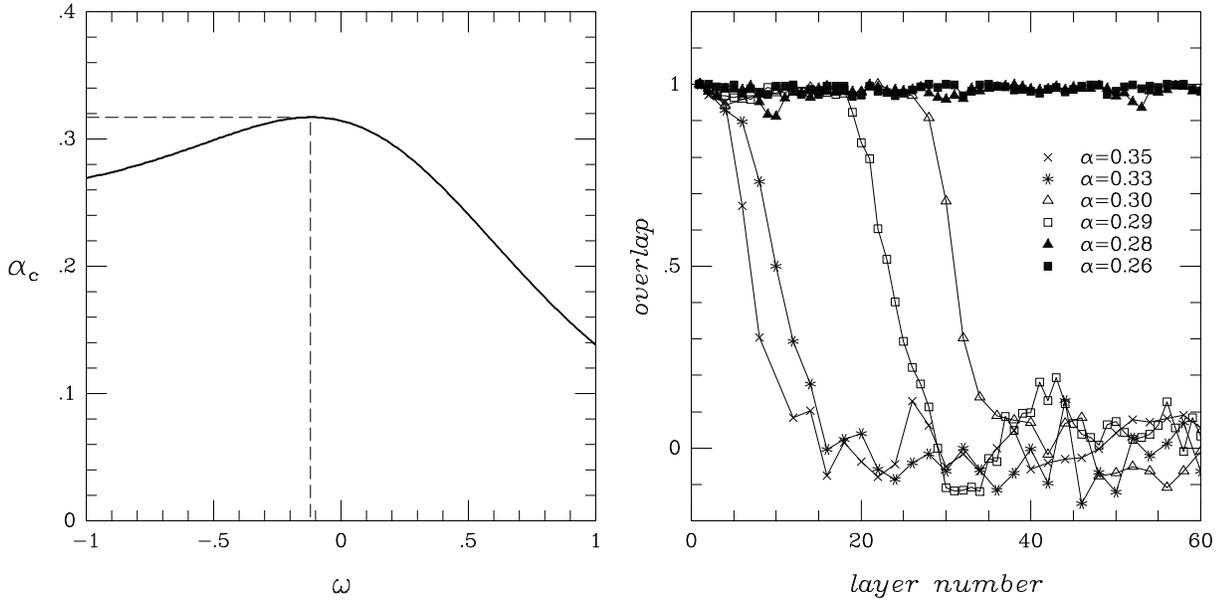}\hspace*{10mm}}
\vspace*{-5mm}
\caption{Left picture - theory: storage capacity $\alpha_c$ for long
chains at $T=0$ (dashed: location of the optimum).   
Right picture - numerical simulations ($N=900$): stationary (pure
state) overlap $m$ as a function
of layer number $\ell$, for  $T=0$ and $\omega=0$. 
}
\label{fig:alphac}
\end{figure}

We first calculate the information storage capacity $\alpha_c$ for
(infinitely) long chains. A 
stationary situation is reached along the chain when 
$(\bm,q,r)=(\bm^\prime,q^\prime,r^\prime)$ 
in (\ref{eq:saddlem}-\ref{eq:saddler}), giving
\be
\bm = \bra \bxi
\int\! Dz~ \tanh\beta\left[\bxi\cdot\bm\plus z \sqrt{\alpha r}\right] 
\ket_{\bxi}
\label{eq:statm}
\ee
\be
q = \bra
\int\! Dz~ \tanh^2\beta \left[\bxi\cdot\bm \plus z \sqrt{\alpha r} \right]
\ket_{\bxi}
\label{eq:statq}
\ee
\be
r=
\frac{(1\minus\omega)^2\plus
q(1\plus\omega)^2\minus 2\omega\beta q(1\plus\omega)(1\minus q)}
{4[1 \minus \frac{1}{2}\beta(1\plus\omega)(1\minus q)]\left[1\minus \beta (1\plus\omega)(1\minus
q)\plus\omega\beta^2(1 \minus q)^2\right]} 
\label{eq:statr}
\ee
Upon taking the limit $T\to 0$ 
and concentrating on 
pure states, where 
$m_\mu=m\delta_{\mu\lambda}$, we can perform the averages
and integrations, 
substitute $x =m/\sqrt{2 \alpha r}$, and simply follow the procedure
described in \cite{amitetal2} to reduce the above set
of equations (\ref{eq:statm}-\ref{eq:statr}) to a single 
transcendental equation (with $m=\erf(x)$):
\be
x\sqrt{2 \alpha}  =
\frac{\erf(x) - \frac{2x}{\sqrt{\pi}} e^{-x^2}}
{\sqrt{\frac{1}{2}(1\plus \omega^2)}} 
\left\{
\frac{\left[ \erf(x) - \frac{2 \omega x}{\sqrt{\pi}} e^{-x^2}\right]
\left[ \erf(x) - \frac{(1+\omega)x}{\sqrt{\pi}}e^{-x^2} \right]}
{\left[ \erf(x) -\frac{2x}{\sqrt{\pi}} e^{-x^2} \right]
\left[ \erf(x) - \frac{\omega^2 + \omega}{\omega^2+1}
\frac{2x}{\sqrt{\pi}} e^{-x^2} \right]}
\right\}^{\frac{1}{2}}
\label{eq:transcendental}
\ee
This equation is to be solved numerically. 
The storage capacity $\alpha_c$ is the value for $\alpha$
for which the nonzero solutions of (\ref{eq:transcendental}) vanish, 
resulting in figure \ref{fig:alphac} (left picture). For 
$\omega=-1$ equation (\ref{eq:transcendental}) reduces
to the  
results of \cite{domanyetal1,domanyetal2,domanyetal3}; for $\omega=1$
it reduces to the results of
\cite{amitetal1,amitetal2}, as it should. 
Somewhat surprisingly, the largest storage capacity 
is obtained for $\omega\sim -0.12$, giving
$\alpha_c\sim 0.317$. We support these analytical 
results with  numerical simulations
on chains with $L=60$ layers of $N=900$ neurons each, for $T=0$ and
$\omega=0$. 
Figure \ref{fig:alphac} (right picture) shows the stationary values
of the condensed overlap
$m$ as a function of the layer number. 
For $\alpha\in \{0.26, 0.28\}$ the desired 
state $m\sim 1$ appears stable, for
$\alpha\in\{0.33,0.35\}$ the state $m\sim 1$ is unstable,
whereas for $\alpha\in\{0.29,0.30\}$ we appear to be close to the critical
value, with $m\sim 1$ appearing stable initially, but eventually
destabilising further down along the chain. This is in reasonable 
agreement with
the theory, which predicts $\alpha_c\sim 0.314$ for
$\omega=0$, if we take finite size effects into account.     

\subsection{Simple Ergodicity Breaking Transitions}
 
Next we turn to transitions  marking the appearance of multiple
(replica symmetric) 
coexistent stable states. We
restrict ourselves to pure states, i.e. in each layer $m_{\mu}= m
\delta_{\mu,\lambda}$ for some $\lambda$ and some layer-dependent $m$, and to the junction between layer
$\ell=1$ and layer $\ell=2$, where the first such transitions are
expected. 
Insertion of pure state solutions, and taking the limit $T\to
0$ in (\ref{eq:saddlem}-\ref{eq:saddler}) again allows us to perform
the averages and integrations. Substitution of 
$y=[J_0m^\prime \plus
Jm]/\sqrt{2 \alpha r^\prime}$ and, in the
case where the first layer evolves freely,  
of $x=J_0m/\sqrt{2 \alpha r}$, 
now leads to the expression $m^\prime=\erf(y)$, in which $y$ is the
solution of the following transcendental equation: 
\be
F(y) = 
y \sqrt{ 2 \alpha} \left[1\plus \rho\left(\frac{1\minus
\omega}{1\plus\omega}\right)^2\right]^\frac{1}{2}\!  
-m\left[\frac{1\minus\omega}{1\plus\omega}\right]
~~~~~~~~~~
F(y)=\erf (y)-\frac{2 y}{\sqrt{\pi}} e^{- y^2}
\label{eq:trans}
\ee
The value of $\rho$ in (\ref{eq:trans}) depends on whether or not
the input layer is clamped into a state, and, in the case of free
relaxation, on the actual value of $m$:
\be
\begin{array}{ll}
{\rm input~clamped:} & m~{\rm given}, ~~\rho=1\\
{\rm free~relaxation:} & m=\erf(x),~~F(x)=x \sqrt{ 2
\alpha}, ~~\rho=[\erf(x)/F(x)]^2\room
\end{array}
\label{eq:inputlayer}
\ee
The simplest situation, $m=0$, already clearly demonstrates  
the distinction between 
the two modes of operation (input clamped versus free relaxation). 
In the clamped case, where $\rho=1$,  
a rescaling of the storage ratio $\alpha$ 
maps (\ref{eq:trans}) onto the equations describing the model of \cite{amitetal1}, and
solutions  with $y\neq 0$ (i.e.
$m^\prime\neq 0$) must therefore bifurcate at 
\be
\alpha_{\rm bif}(\omega)=
\frac{\alpha_{\rm bif}(1)}{\sqrt{1\plus
\left(\frac{1- \omega}{1+\omega}\right)^2}}
\sim 
\frac{0.138}{\sqrt{1\plus
\left(\frac{1- \omega}{1+\omega}\right)^2}}
\ee
On the other hand, for free relaxation of the first layer we find
$\rho=\infty$, so $y=0$ (i.e. $m^\prime=0$ is the only solution of
(\ref{eq:trans})). In the first case the imposed $m=0$ state of layer one is like a
paramagnetic state; in the second case the $m=0$ state is of a spin-glass
type. \vsp
 
\begin{figure}[t]
\centering
\vspace*{160mm}
\hbox to \hsize{\hspace*{-5mm}\includegraphics{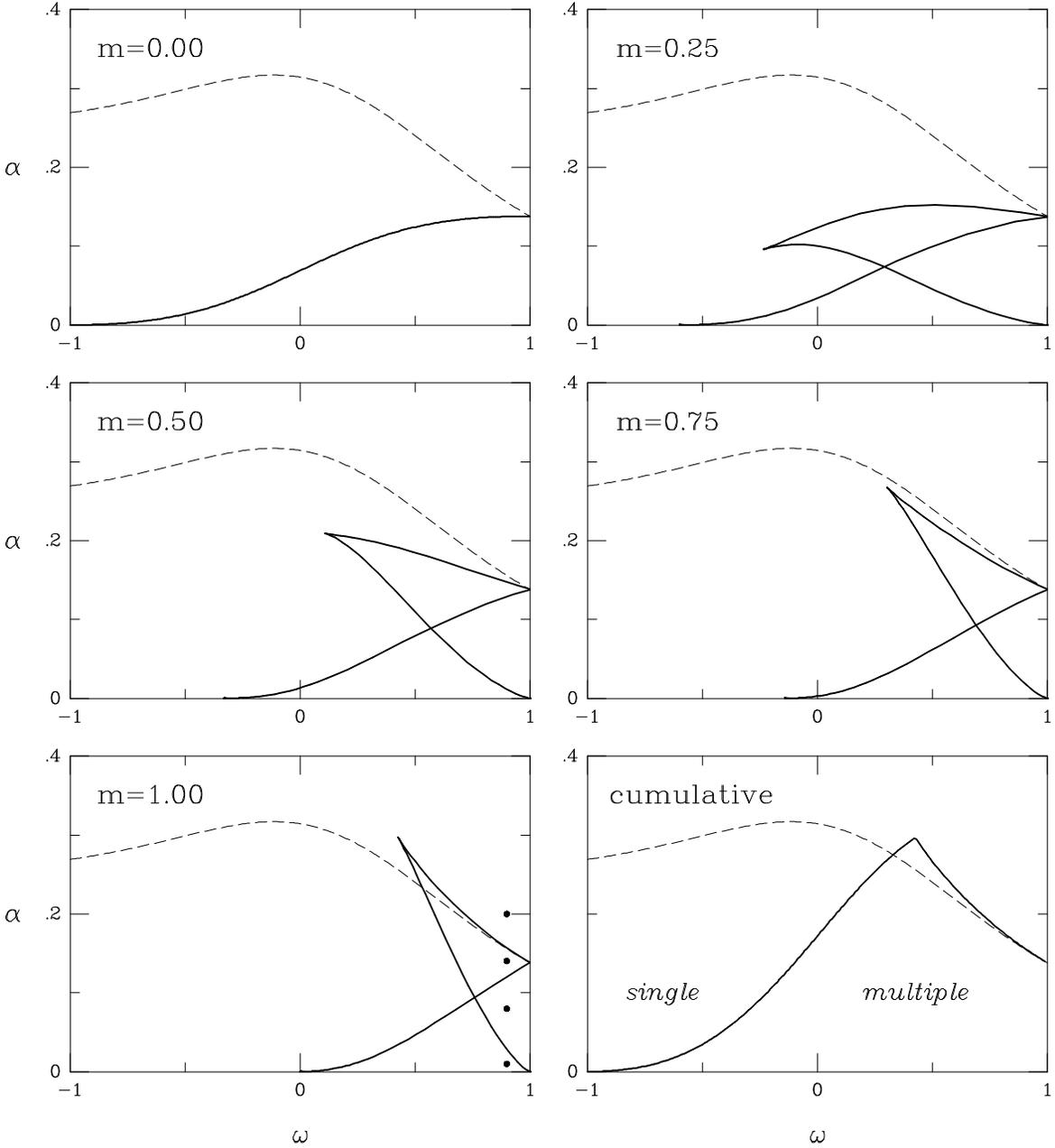}\hspace*{5mm}}
\vspace*{-5mm}
\caption{Examples of bifurcation lines, marking the appearance of
multiple pure (and replica-symmetric) stable states in layer 2, for clamped
operation. 
Left of and above the solid lines: a single stable state; close to the $\omega=1$ line: multiple
stable states. Dashed lines: the long-chain storage capacity
$\alpha_c$. 
Dots in $m=1.00$ picture: 
control parameters $(\omega,\alpha)$ used in numerical simulations (described
below).  Bottom-right  picture: boundary of the cumulative region 
where multiple stable states
exist (obtained by combining the
results for input overlaps $m\in[0,1]$).}
\label{fig:bifs}
\end{figure}

\begin{figure}[t]
\centering
\vspace*{120mm}
\hbox to \hsize{\hspace*{-5mm}\includegraphics{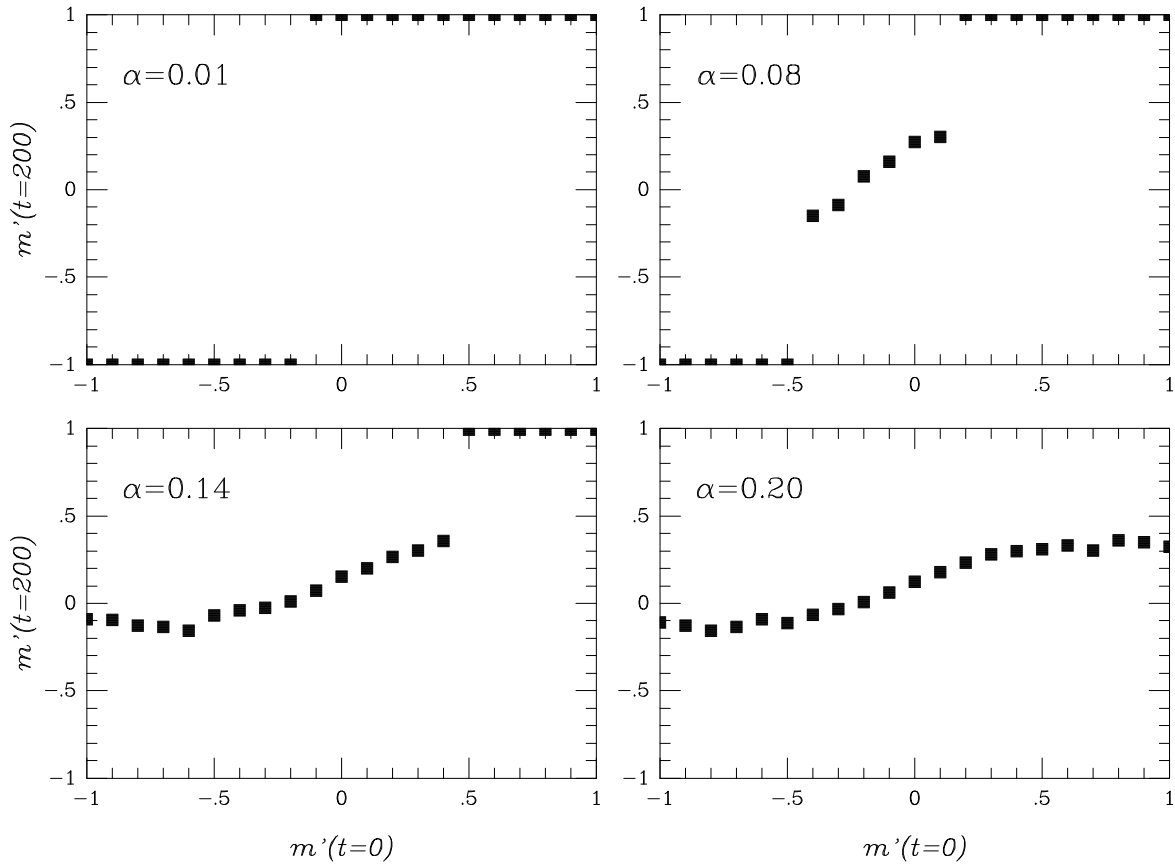}\hspace*{5mm}}
\vspace*{-5mm}
\caption{Simulation results ($N=12,000$), indicating the number of  
(pure) stable states in layer 2.  Each picture shows
the overlap $m^\prime$ at time $t=200$
versus the overlap $m^\prime$ at $t=0$, for $\omega=0.9$ and input
overlap $m=1$. 
The $\alpha$ values correspond to the dots in figure \ref{fig:bifs}.
}
\label{fig:ergos}
\end{figure}

For arbitrary $m$ we obtain the condition for new solutions of
(\ref{eq:trans}) to bifurcate by derivation of (\ref{eq:trans}) with
respect to $y$. This gives a new equation, to be solved simultaneously
with (\ref{eq:trans}); a simple transformation allows us to rewrite
the resulting pair as
\be
\alpha = \frac{8}{\pi} y^4 e^{-2y^2}
\left[1\plus \rho\left(\frac{1\minus
\omega}{1\plus\omega}\right)^2\right]^{-1}
~~~~~~~~~~
m\left[\frac{1\minus\omega}{1\plus\omega}\right]=G(y)
\label{eq:bifurcations}
\ee
(to be solved with the appropriate values for 
$(m,\rho)$, given in (\ref{eq:inputlayer})), and with 
\be
G(y)= \frac{2 y}{\sqrt{\pi}}(1\plus 2y^2)  e^{-y^2} 
- \erf (y)
\ee
It is clear from (\ref{eq:bifurcations}) that
bifurcations of new solutions can only occur for sufficiently small
$\alpha$.   
For clamped input operation, where the input overlap
$m$ is an independent parameter,  
numerical solution of (\ref{eq:bifurcations}) results in bifurcation
lines in the $(\omega,\alpha)$ plane, shown in figure
\ref{fig:bifs}. 
Left and above the solid lines
are the regions with a unique stable state, at the other side of the
solid lines (near $\omega=1$) multiple stable states
exist. The bottom-right picture shows the result of combining all regions with
multiple stable states, for $m\in[0,1]$: left of the solid line there
is a single stable state in the second layer, 
irrespective of $m$, right of the solid line input
overlaps  $m\in[0,1]$ can be found which give rise to multiple stable states
in the second layer.    
These results are again supported by numerical simulations: figure
\ref{fig:ergos} shows the relation between initial and (almost) 
\clearpage
\noindent final
overlaps in layer 2, for $N=12,000$ and $m=1$,  
and the four $(\omega,\alpha)$
combinations indicated with dots in figure \ref{fig:bifs}. The number
of stable states predicted 
(figure \ref{fig:bifs}) are $(2,3,2,1)$, for the four cases
$\alpha=(0.01,0.08,0.14,0.20)$, respectively. This is in perfect
agreement with the simulation results of figure
\ref{fig:ergos}, where for each graph 
the number of
stable states is the number of discontinuities plus one. For a system
in equilibrium one expects horizontal line segments between  
the discontinuities; apparently for $\alpha\in\{0.08, 0.14, 0.20\}$ the state at $t=200$ is not an
equilibrium state yet (due to the large
relaxation times involved).     
 
For free relaxation of the  first layer,     
a nonzero $m$, being  a
solution of (\ref{eq:inputlayer}), is a monotonically decreasing 
function of $\alpha$ with
$m(\alpha)\geq 0.966$.    
The bifurcation lines obtained by solving numerically
(\ref{eq:inputlayer},\ref{eq:bifurcations}) 
are now found to be practically indistinguishable from those
of the clamped case with $m = 1$ (see figure \ref{fig:bifs}), 
provided $\alpha \leq
\alpha_c(\omega=1)\sim 0.138$. For $\alpha>\alpha_c(\omega=1)$ we are
back at $m=0$, where  no bifurcations were found to be allowed. 
  
\subsection{The AT Instability}

The third and final type of transition to be analysed is the 
AT-line \cite{AT}, where the relevant saddle-point of (\ref{eq:RSB}) ceases to
be replica-symmetric. 
Note that the fully recurrent limit $\omega\to 1$ of the present model
exhibits broken
replica symmetry (RSB) near $T=0$ \cite{amitetal1}, whereas 
the  $\omega\to -1$ limit
does not \cite{domanyetal1}, which hints at the
possible existence of an RSB transition at some critical value 
$\minus 1< \omega_{\rm AT}<1$ for $T=0$. 
In order to perform a bifurcation analyis {\em a la} \cite{AT}, we go 
back to equation (\ref{eq:RSB}), 
and work out  the RSB saddle point equations:
\bea
m_\alpha &=&
\bigbra\frac{\sum_{\bsigma}~ \sigma_\alpha ~e^{\beta
\suma \sigma^\alpha\sum_{\mu\leq k} [J_0 m_\mu^\alpha +
J \tilde{m}_\mu]\xi_\mu - i\sum_{\alpha\beta}\sigma_\alpha\hq_{\alpha\beta}
\sigma_\beta}}
{\sum_{\bsigma} e^{\beta
\suma \sigma^\alpha\sum_{\mu\leq k} [J_0 m_\mu^\alpha +
J \tilde{m}_\mu]\xi_\mu - i\sum_{\alpha\beta}\sigma_\alpha\hq_{\alpha\beta}
\sigma_\beta}}\bigket_\bxi\\
q_{\alpha \beta} &=&
\bigbra\frac{\sum_{\bsigma}~ \sigma_\alpha\sigma_\beta ~e^{\beta
\suma \sigma^\alpha\sum_{\mu\leq k} [J_0 m_\mu^\alpha +
J \tilde{m}_\mu]\xi_\mu - i\sum_{\alpha\beta}\sigma_\alpha\hq_{\alpha\beta}
\sigma_\beta}}
{\sum_{\bsigma} e^{\beta
\suma \sigma^\alpha\sum_{\mu\leq k} [J_0 m_\mu^\alpha +
J \tilde{m}_\mu]\xi_\mu - i\sum_{\alpha\beta}\sigma_\alpha\hq_{\alpha\beta}
\sigma_\beta}}\bigket_\bxi
\label{eq:RSBq}\\
\hatq_{\alpha \beta} &=& \frac{1}{2} i \beta^2 J^2 x_\alpha x_\beta
[\sum_{\mu > k}\tilde{m}_\mu^2]
+ \frac{1}{2} i \alpha \beta J_0
\frac{ \int\! D\bz ~ z_\alpha z_\beta~
e^{\frac{1}{2} \beta J_0 \bz\cdot \bq\bz}}
{ \int\! D\bz ~
e^{\frac{1}{2} \beta J_0 \bz\cdot \bq\bz}}
\label{eq:RSBqhat}
\eea
with $\bsigma\in\{\minus 1,1\}^n$, $\bz\in\Re^n$,
$D\bz=(2\pi)^{-n/2}e^{-\frac{1}{2}\bz^2}d\bz$,  and 
$x_\alpha = \sum_\gamma \left[ \one \minus \beta
J_0\bq\right]^{-1}_{\alpha \gamma}$. 
\begin{figure}[t]
\centering
\vspace*{78mm}
\hbox to \hsize{\hspace*{20mm}\includegraphics{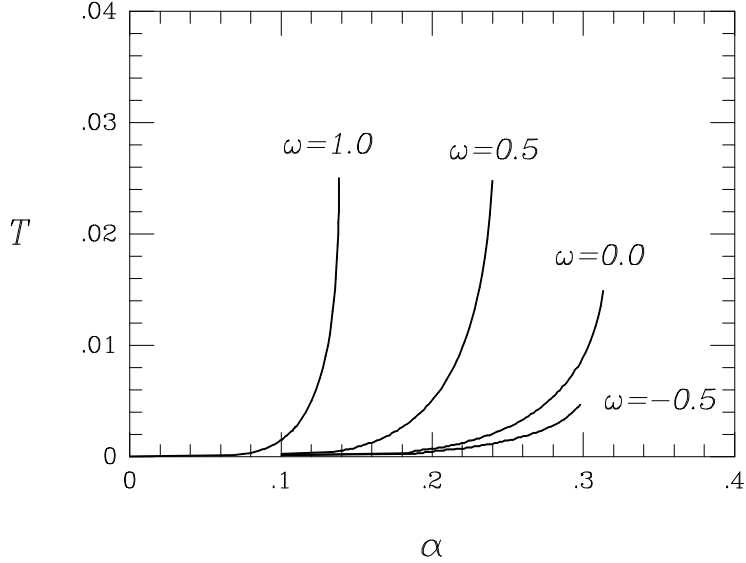}\hspace*{-20mm}}
\vspace*{-3mm}
\caption{Location of the continuous replica symmetry breaking
transition (AT line) in
the $(\alpha,T)$ plane. For $\omega=-1$ the AT line collapses onto the
line $T=0$.  }
\label{fig:atline}
\end{figure} 
\noindent Following \cite{AT} 
we now consider the so-called replicon fluctuations around the RS solution (\ref{eq:saddlem},\ref{eq:saddleq},\ref{eq:saddler}):
\be
\qab\to \delta_{\alpha \beta}\plus 
q(1\minus\delta_{\alpha \beta})\plus \delta q_{\alpha \beta}~~~~~~~~
\hat{q}_{\alpha\beta}\to \frac{1}{2}i\alpha\beta^2[R\delta_{\alpha \beta}\plus 
r(1\minus\delta_{\alpha \beta})]\plus \delta \hat{q}_{\alpha \beta}
\label{eq:replicon}
\ee
with the
properties  
$|\delta q_{\alpha \beta}| \ll 1$,
$\delta q_{\alpha \beta} = \delta q_{\beta \alpha}$,
$\suma \delta q_{\alpha \beta} = 0$, and  
$\delta q_{\alpha \alpha} = 0$. 
Working out the (coupled) variations in
(\ref{eq:RSBq},\ref{eq:RSBqhat}) due to (\ref{eq:replicon}), and
requiring such variations to lead to  an instability for $n\to 0$ (a
massless mode in first order perturbation theory),  
gives, after a modest amount of algabra,  the condition
\be
1=\frac{\alpha(\beta J_0)^2}{[1\minus\beta J_0(1\minus
q)]^2}\bra\int\!Dz~\cosh^{-4}\beta\left[\bxi\cdot(J_0\bm\plus
J\tilde{\bm})\plus z\sqrt{\alpha r}\right]\ket_\bxi 
\label{eq:atline}
\ee
The RS solution (\ref{eq:saddlem},\ref{eq:saddleq},\ref{eq:saddler}) 
is stable if the right-hand side of (\ref{eq:atline})
is smaller than one. For $J_0=0$ (i.e. for the feed-forward model of
\cite{domanyetal1,domanyetal2,domanyetal3}) clearly no RSB occurs. 
For $J_0>0$ (i.e. for $\omega>\minus 1$) we
have to solve (\ref{eq:atline}) numerically, in combination with the
RS saddle-point equations
(\ref{eq:saddlem},\ref{eq:saddleq},\ref{eq:saddler}).     
For stationary states in large chains (i.e.
$(\bm,q,r,)=(\bm^\prime,q^\prime,r^\prime)$
and $L\to \infty$),
and  pure states $m_\mu=\delta_{\mu\lambda}$, the solution thus obtained is
shown in figure \ref{fig:atline} for $\omega\in\{-0.5,0,0.5,1.0\}$ 
(for $\omega=-1$ the AT line collapses onto the line $T=0$). 
Our first conclusion is that for $T=0$ the RSB transition occurs at
$\omega=-1$; any finite fraction of recurrence in the interactions
apparently destabilises the replica symmetric solution. 
Secondly, although we
cannot interpret the $T>0$ data  in figure \ref{fig:atline} directly in terms
of the operation of the present model, the fact that,  
as we move away from the fully recurrent case $\omega=1$, the noise
level $T$ at which replica symmetry breaks decreases monotonically,
suggests that we can
be confident that for the present model, as for the fully recurrent
case, replica symmetry breaking
effects will be of a minor quantitative nature only. This, in fact, is
borne out by the agreement observed between simulations and the RS
solution. 

\section{Discussion}

In this paper we have analysed a simple 
layered Ising spin neural network model, with   
a controllable competition between 
recurrent and feed-forward information processing. This model interpolates  
between the fully recurrent and symmetric attractor model of 
\cite{amitetal1,amitetal2} and the strictly feed-forward model of
\cite{domanyetal1,domanyetal2,domanyetal3}. 
At zero noise level and in a stationary state, the model 
can be solved analytically near
saturation, using replica theory (where we have restricted ourselves
to the replica-symmetric ansatz),  in spite of its interaction matrix 
being neither 
symmetric nor strictly feed-forward (which are the features on which
analysis usually relies). 
This property also turns it into a nice 
toy model for testing new tools for tackling the stationary
states  of neural network models without detailed balance. 

In  two extreme limits,  
fully recurrent and fully
feed-forward operation, respectively, the results of 
\cite{amitetal1,amitetal2} and
\cite{domanyetal1,domanyetal2,domanyetal3} 
are recovered correctly, as it should.  
In the intermediate regime, the built-in competition between 
recurrent operation (which is highly
non-ergodic within individual layers) versus feed-forward operation
(which is ergodic
within layers), is reflected in a non-trivial way in various types
of transitions. These describe saturation breakdown, 
simple ergodicity-breaking involving pure states, and the AT
instability \cite{AT} with respect to replic-symmetry-breaking.  
The largest storage capacity is found to be $\alpha_c\sim 0.317$, 
which is obtained for a specific balance of the two types of
interaction.  Replica symmetry turns out to break down as soon as one 
moves away from the strictly feed-forward limit,
i.e. for  any finite fraction of recurrence in the
interactions. 

Our results, which
are supported by numerical simulations, 
might also play a role in describing 
the competition between recurrent and feed-forward operation 
in the peripherical regions of the brain (upon suitable quantitative 
adaptation of model details), where architectures similar to the one
studied in this paper can be found.

\subsubsection*{Acknowledgements}

The authors wish to thank the National University of
Mexico (UNAM) (DGAPA Project IN100294) 
and the J.S. McDonnel 
Foundation (Visiting Fellowship, LV)  for support.

\end{document}